# The Effect of Charge Carrier Cooling on the Ultrafast Carrier Dynamics in $Cs_2AgBiBr_6$ Thin Films

Huygen J. Jöbsis[1,2], Lei Gao[3], Antti-Pekka M. Reponen[2], Zachary A. VanOrman[2], Rick P.P.P.M. Rijpers[1], Hai I. Wang[1,3], Sascha Feldmann[2,4], Eline M. Hutter*[1]

1) Utrecht University, Princetonlaan 8, 3584 CB Utrecht, The Netherlands

2) Rowland Institute, Harvard University, 100 Edwin H Land Blvd, Cambridge, MA 02142, USA

3) Max Planck Institute for Polymer Research, Ackermannweg 10, 55128 Mainz, Germany

4) Institute of Chemical Sciences and Engineering, École Polytechnique Fédérale de Lausanne, Rue de l'industrie 17, 1951 Sion, Switzerland

*Corresponding author: e.m.hutter@uu.nl

**Abstract**

Cs$_2$AgBiBr$_6$ shows promise for solution-processable optoelectronics, such as photovoltaics, photocatalysis and X-ray detection. However, various spectroscopic studies report rapid charge carrier mobility loss in the first picosecond after photoexcitation, limiting carrier collection efficiencies. The origin of this rapid mobility loss is still unclear. Here, we directly compare hot excitation with excitation over the indirect fundamental bandgap, using transient absorption and THz spectroscopy on the same Cs$_2$AgBiBr$_6$ thin film sample. From transient absorption spectroscopy, we find that hot carriers cool towards the band-edges with a cooling rate of 0.58 ps$^{-1}$, which coincides with the observed mobility loss rate from THz spectroscopy. Hence, our study establishes a direct link between the hot carrier cooling and ultrafast mobility loss on the picosecond timescale.

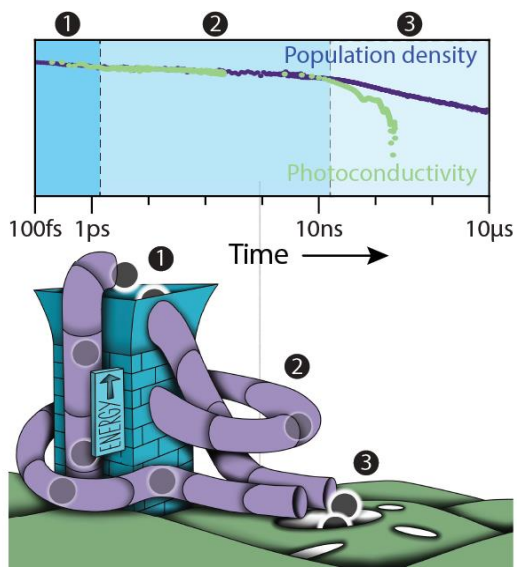

Halide-elpasolites have emerged as a promising semiconductor material class for light conversion applications, due to their tunable optical properties, chemical stability, solution processability, and photostability.[1,2] This versatility is reflected in the materials crystal structure, $A_2BB'X_6$, consisting of (at least) four different ions, opening up an immense number of potential compositions and thus material properties.[3,4] Typically, two corner-sharing metal-halide octahedra ($[B(I)X_3]^{2-}$, $[B'(III)X_3]$) alternate each other in a cubic crystal lattice with monovalent A-site cation occupying the interstitial sites between the octahedra. Control over the optical and mechanical properties can be achieved by varying the chemical composition. Typically, substitution of the halides with larger halides reduces the bandgap energy[5–7] and softens the crystal lattice.[8] Additionally, exchange of the B'(III)-site cation provides control over the nature of the bandgap, and the bandgap energy.[9–11] Silver-bismuth halide-elpasolites, and particularly $Cs_2AgBiBr_6$, are the most studied compositions due to their excellent stability, sizeable bandgap in the visible part of the spectrum, and relatively benign heavy-metals. As such, numerous demonstrations of $Cs_2AgBiBr_6$-based applications can be found in the literature, including X-ray detection, photochemical reactions, humidity sensing, photovoltaics, and memory storage.[12–17]

The photoconversion performance of $Cs_2AgBiBr_6$, however, is lagging behind compared to state-of-the-art halide-perovskite semiconductors. Most likely this is due to the rapid loss of photoconductivity observed on a picosecond timescale.[18–24] In addition, the photoluminescence (PL) energy has been reported to be roughly 500 meV below the optical absorption onset.[25] The loss mechanism(s) behind the mobility loss of charge carriers and the strongly red-shifted PL is still under debate. One hypothesis is the formation of a polaron state due to strong charge carrier–phonon coupling, restricting or localizing the charge state down to a single unit cell.[18,22,26,27] Another interpretation of such a localized state is the formation of a self-trapped exciton due to local deformations of the crystal lattice upon photoexcitation, which could explain the red-shifted PL signal as well.[21] Except for the work by Zhang et. al.,[23] most previous reports make use of excitation energies that are absorbed by the strong direct transition at 2.8 eV. The fundamental bandgap of $Cs_2AgBiBr_6$ on the other hand is indirect, and has a value of ~2.2 eV. As such, especially on the sub-picosecond timescales, the decay dynamics are strongly affected by cooling of charge carrier in both energy and *k*-space. So, careful analysis is required to disentangle cooling effects from losses in photo-induced carrier population and mobility, ideally on the same samples.

Here, we perform multi-resonant-pump broadband visible- and terahertz-probe spectroscopy experiments to elucidate the ultrafast charge carrier dynamics in $Cs_2AgBiBr_6$ thin films. By comparing transient absorption (TA) spectroscopy measurements at high-energy excitation with band-edge excitation, we are able to isolate the ultrafast hot carrier cooling dynamics and extract a cooling rate constant of ca. 0.58 ps$^{-1}$ (using a pump fluence of $2.5 \times 10^{13}$ cm$^{-2}$). The observed cooling rate coincides with the mobility loss observed with optical-pump terahertz-probe (OPTP) experiments using identical excitation energies. Based on these results we conclude that the rapid decay dynamics observed on a sub-picosecond timescale can be solely explained by the cooling of hot charge carriers, and that the hot carriers are more conductive than the cold ones. We note that the increased conductivity of hot carriers has been observed before for this material,[23] and that its origin remains unsolved. Our work provides a rationale behind the ultrafast dynamics in $Cs_2AgBiBr_6$ thin films.

Thin layers of $Cs_2AgBiBr_6$ were spin-coated on quartz glass substrates following a synthesis route adapted from literature,[28] see Experimental Methods in the Supporting Information (SI). The cubic crystal structure is confirmed using X-ray diffraction crystallography and the layer thickness is estimated to be 220 nm using our previously reported extinction coefficient, $\kappa$,[29] and the transmittance, $T$, spectrum (**Fig. S1**). The steady-state optical properties are shown in **Figure 1a** displaying the characteristic direct transition absorption (dip in $T$) around 2.8 eV. This pronounced feature originates from a direct transition at the $\Gamma$-point ($E_{dir}$), which is above the fundamental bandgap transition of this material (**Fig. 1b**).[5,30] The indirect bandgap transition ($\Gamma$ to $L$, $E_{in}$) gives rise to the gradual absorption onset around 2.2 eV. The indirect bandgap absorption is strongly red-shifted with respect to the direct excitonic transition. This spectral separation provides a unique situation where bandgap excitation is expected to result in the bleaching of distinct spectral regions in TA spectroscopy, corresponding to the direct and indirect transitions. That is, excitation into the indirect bandgap at 2.2 eV (green arrow in **Fig. 1b**), populates holes in the valence band maximum (VBM) at $\Gamma$, and electrons in the conduction band minimum (CBM) around the $L$ valley. The VBM hole population at $\Gamma$ leads to bleaching of the direct transition at 2.8 eV. In the case of exciting into the direct transition (blue arrow in **Fig. 1b**), hot electrons at $\Gamma$ initially also contribute to the bleach at 2.8 eV. The hot electrons cool in energy and in $k$-space, and no longer contribute to the direct bleach (at 2.8 eV) when residing at the band-edge at $L$. As a consequence, comparing blue (hot) excitation with green (band-edge) excitation, directly probes the cooling from $\Gamma$ to L. To probe the hot carrier cooling mechanism we use ultrafast TA spectroscopy on a sub-picosecond timescale to study the population dynamics of excited states ($N(t)$). On the other hand, OPTP complementarily probes the conductivity of all mobile carriers independent of their position in $k$-space, i.e. the transient product of charge carrier population and mobility ($N(t) \times \mu(t)$). Therefore, combining these two techniques at different excitation energies allows us to distinguish ultrafast phenomena such as mobility loss and population decay *via* hot carrier cooling.

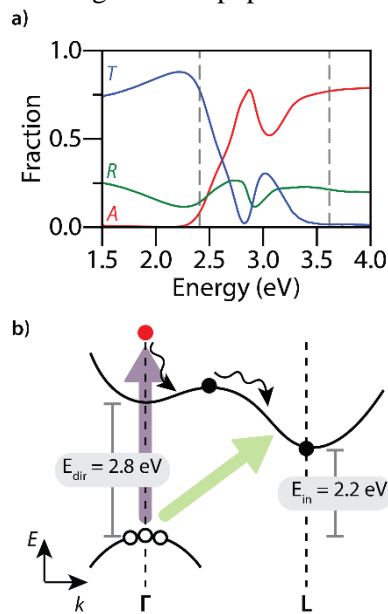

**Figure 1** – Optical properties of $Cs_2AgBiBr_6$ thin films. **a)** Steady-state optical properties, absorptance ($A$, red), transmittance ($T$, blue) and reflectance ($R$, green) of a 220 nm thin $Cs_2AgBiBr_6$ film on a glass substrate. **b)** Schematic representation of the band structure of $Cs_2AgBiBr_6$ showing the direct bandgap $\Gamma$ to $\Gamma$ transition and the indirect $\Gamma$ to $L$ transition. The blue and green arrows represent the pump energies used in the TA and OPTP experiments, 343 nm (3.6 eV) and 515 nm (2.4 eV), respectively which are indicated with grey dashed lines in **a**).

TA spectroscopy is a pump–probe technique where a broadband probe-pulse is used to record the transmission spectrum before and after exciting the sample with a narrow-band laser pulse, the pump. Upon excitation, electrons and holes populate excited states temporally changing the dielectric function of the material. In the low excitation limit, the change in the dielectric function scales with the excited-state population density. As such, the difference spectrum of the excited-state and steady-state transmission provides insight in the photoexcited charge carrier population. By varying the delay time, $t$ — the arrival time difference between pump and probe (which records the excited state spectrum) — the decay dynamics of the excited-state population can be traced. **Figure 2a** shows the difference transmittance spectra normalized for the steady-state transmittance ($\Delta T/T$) of Cs$_2$AgBiBr$_6$ at $t$ = 300 fs, when exciting the sample at 343 nm (ca. 3.6 eV, blue arrow in **Fig. 1a**) and 515 nm (ca. 2.4 eV, green arrow in **Fig. 1a**) at similar excitation densities (see Supplementary note **S1** and **Fig. S2**). The blue pump-pulse excites charge carriers well above the direct bandgap, creating hot carriers. These hot carriers thermalize and relax back to the band edges typically within a few picoseconds.[31–36] By exciting just over the indirect bandgap using a 515 nm pump-pulse, the charge carriers immediately reside at the band edges of $\Gamma$ (hole) and $L$ (electron), respectively. As the probability for this electronic transition is much smaller than that for the direct transition, we also observe weaker signals in TA. Importantly, as $E_{in} \ll E_{dir}$, the observed bleach signal at 2.8 eV reports only the hole population at the valence band maximum by 515 nm (2.4 eV) excitations (**Fig. 1b**). Therefore, the difference trace of this direct $\Gamma$-to-$\Gamma$ transition for hot (343 nm) and indirect bandgap (515 nm) excitation reveals the electron cooling dynamics (black curve in **Fig. 2c**). The hot carrier population decays monoexponentially with a cooling rate of 0.58 ps$^{-1}$ (using a pump fluence of 2.5×10$^{13}$ cm$^{-2}$). This is supported by following the position of the $\Delta T/T$ maximum around 2.8 eV over time (**Fig. 2d**). In the first picosecond after photoexcitation with the 343 nm pump, the bleach maximum around 2.8 eV is red-shifted. A red-shift of the resonance energy is typically observed for carrier cooling as the lattice temperature is increased.[37,38] On the other hand, for $t < 1$ ps, the bleach position blue-shifts when exciting just over the bandgap, which is a result of state-filling (Pauli-blocking).[39]

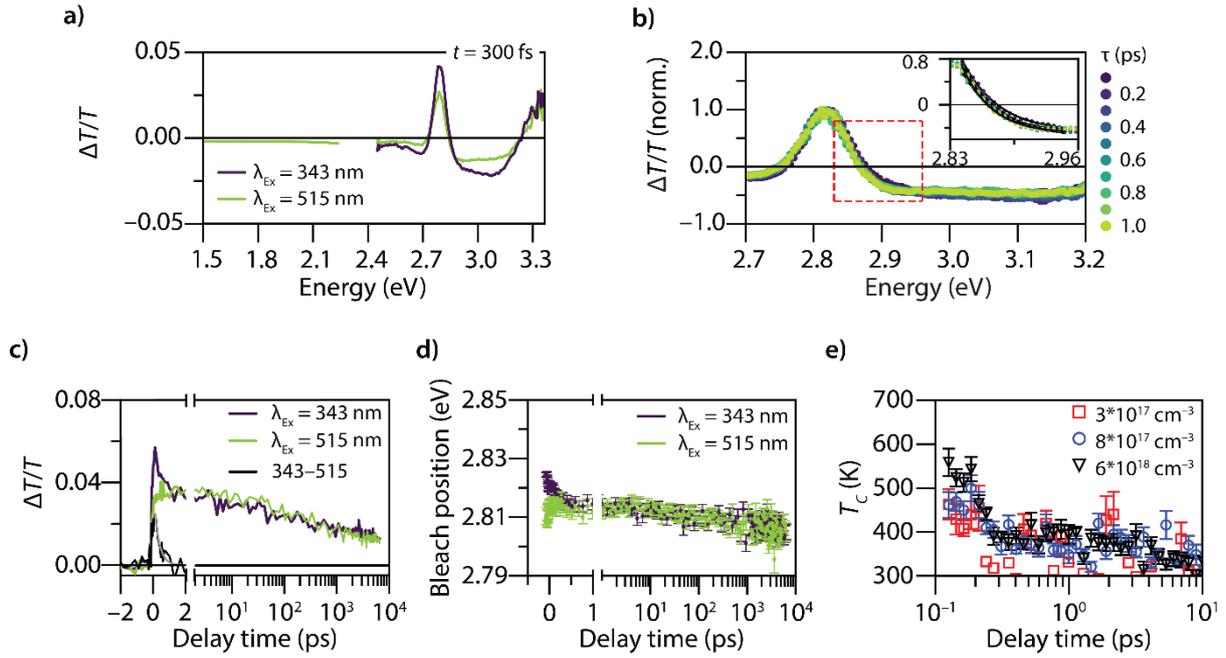

**Figure 2** – Transient absorption decay dynamics of photoexcited electron and holes. Transient transmittance $\Delta T/T$, spectral slice after 300 fs for excitation at 343 nm (3.6 eV, blue) and 515 nm (2.4 eV, green). **b)** The presence of hot carriers broadens the bleach feature on the high energy side. The zoom in displays the Maxwell–Boltzmann distributions (black) fitted to extract the hot carrier temperature. **c)** Recovery time trace of the direct transition (2.78 – 2.85 eV) representing the electron and hole population (hot excitation at 343 nm, blue trace, excitation density 8.33 × 10$^{17}$ cm$^{-3}$) and just the hole population (band-edge excitation at 515 nm, green trace, excitation density 1.03 × 10$^{18}$ cm$^{-3}$). The difference trace (343 − 515, black) is used to determine carrier cooling from $\Gamma$ to $L$. **d)** Bleach maximum position of the direct transition as a function of delay time for excitation at 343 nm (blue) and 515 nm (green). **e)** Extracted $T_c$ as a function of delay time for excitation different excitation densities.

Spectrally the presence of hot carriers broadens the bleach feature on the high-energy side as the photo-induced transient carrier temperature, $T_C$, is much higher than the lattice temperature (**Fig. 2b**). The photo-induced carrier temperature can be extracted using the Maxwell–Boltzmann approximation of the Fermi-Dirac distribution function.[33–35,40–42] We are aware that this methods relies on an arbitrarily chosen fitting region,[43] however, as demonstrated in the Supplementary Information different fitting ranges result in similar $T_C$ values and cooling rates for this material (Supplementary note **S2** and **Fig. S3**). **Figure 2e** shows the extracted photo-induced carrier temperature as a function of time for different pump-energies. These temperatures are in line with other semiconductors with similar bandgap energy and experimental conditions.[40,44]

To study the temperature-dependent carrier cooling dynamics, we perform transient transmittance experiments between 10 and 300 K, again using either 343 nm or 515 nm pump. At lower temperatures following 343 nm excitation, the hot carrier population at a given delay (e.g. for $t = 1$ ps) is expected to increase due to reduced phonon population for cooling and is experimentally observed (**Fig. 3a**).

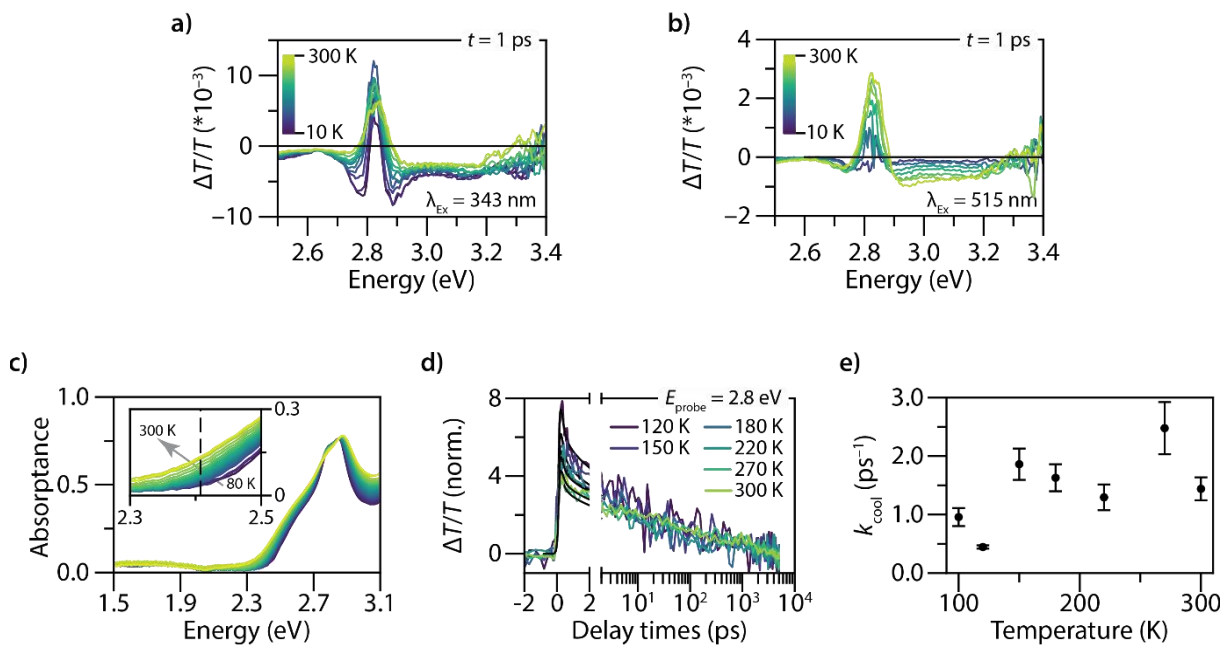

**Figure 3** – Temperature-dependent optical properties. $\Delta T/T$ spectral slices (at $t = 1$ ps) between 10 – 300 K for **a)** 343 nm and **b)** 515 nm excitation. **c)** Temperature-dependent linear $A$ between 80 – 300 K. Zoom-in shows $A$ around the 515 nm pump energy. **d)** Temperature-dependent recovery traces of the direct bleach transition when exciting with 343 nm. The traces are normalized to unity at 1 ns delay times. **e)** Hot carrier cooling rate, $k_{cool}$, as a function of temperature.

increases as the amplitude of the recovery traces increases when pumping at 343 nm. In contrast, for bandgap excitation (515 nm pump) the TA signal gradually decreases upon cooling, with virtually no TA signal left at 10 K (**Fig. 3b**). This can be understood as the phonons required for this indirect transition are suppressed at lower temperatures, also thus confirming experimentally again its indirectness in nature. We further confirm this statement by temperature-dependent optical absorption spectroscopy shown in Figure 3c: we observe a slight blue shift of the absorption onset resulting in much reduced optical absorption (and thus the bleach signal) at 515 nm excitation at lower temperatures.

Similar behavior has been observed previously for other semiconductors such as Si, GaAs, InP and InAs by Varhsni (**Fig. S4**).[45] Moreover, we observe a small hysteresis in the absorption onset energy around 100 K (**Fig. S4**) over a temperature range of 20 K. Around these temperatures the crystal lattice undergoes a phase transition from a cubic to a tetragonal phase.[46] For excitation at 343 nm the hot carrier population seems to increase, as $\Delta T/T$ increases upon lowering the temperature to 120 K. This cannot be explained by an increase in excitation density, as the transmittance at 3.6 eV (343 nm) is practically zero at room temperature (**Fig. 1a**). Therefore the absorptance cannot further increase upon lowering the temperature. Despite the increased carrier population and suppression of phonon vibrations at low temperature the cooling rate, $k_{cool}$, remains constant as a function of temperature (**Fig. 3d, e**). This was similarly observed for the carrier localization rate, assigned by previous OPTP experiments.[22] We note that the determined cooling rates as a function of temperature are higher than those we observed in **Figure 2** (for similar pump fluences). We attribute this to the lower time resolution due to additional chirp from the cryostat windows (**Fig. S5**) in our measurements. The similar cooling dynamics compared to the excited-state localization dynamics reported in literature suggest that the observed mobility loss could in principle be explained by a decreased population density of mobile hot carriers. To investigate the photoconductivity dynamics we perform OPTP measurements on the same sample using similar pump energies and fluences.

The electric field applied by the terahertz probe (~meV) interacts with mobile photoexcited carriers, thus probing the photoconductivity of the sample. The photoconductivity is proportional to the product of the charge carrier mobility and the carrier density. Similar to previous reports, the photoconductance gradually increases in the first 2 ps to reach a plateau after excitation into the indirect bandgap, whereas a prompt increase of the photoconductance followed by a fast picosecond decay is observed for hot excitations (**Fig. 4a**).[23] Note that the photoconductivity traces in **Figure 4a** are divided to the excitation density. So, the overlapping traces for $t > 2ps$ indicate that the same final state is formed with the same yield mobility product. Similarly overlapping decay dynamics are observed for the population density probed with the visible probe in TA-experiments (**Fig. 2c**). Comparison of the TA recovery trace for bandgap excitation— that scales with the photoexcited hole population — with the OPTP time trace reveal strikingly similar dynamics (**Fig. 4b**). This indicates a constant mobility of the band edge charge carriers, of ca. 2 cm$^2$ V$^{-1}$ s$^{-1}$ (**Fig. S6**). Interestingly, for hot excitation a strong overlap of the TA and OPTP recovery traces is observed as well (**Fig. 4c**). This overlap, for $t > 2$ ps, suggests that the charge carriers at the band edge have a constant mobility independent of the excitation energy and that the hot carrier mobility is considerably increased. So, in line with previous reports in the literature,[23] we attribute the increased photoconductivity to the presence of mobile hot carriers, with a yield mobility product roughly 4 cm$^2$ V$^{-1}$ s$^{-1}$. The origin of such increased conductivity of hot carriers remains unsolved and is in contrast to other semiconductors (e.g. silicon).[47,23] Previous temperature-dependent mobility measurements found that the mobility of long-lived species increases with temperature, consistent with small polaron behaviour.[22,23] At the same time, the short-lived peak photoconductivity exhibits band-like transport, with its amplitude decreasing as temperature rises, reflecting that hot carriers display delocalized transport. This delocalization-to-localization transition is proposed to originate from a large-to-small polaron transformation.[48] To avoid confusion, we will continue to refer to these species as polarons. Furthermore, as discussed in the introduction, the polaronic nature of final charge states have been reported in previous work, but whether these states are "excitonic" or "single-charge" states remains controversial. For that, we consider the complex photoconductivity recorded in our THz spectroscopy experiments. If excitons were formed, one would expect a reduction in the real part of the photoconductivity, as the population of free conductive carriers would decrease. Additionally, excitons typically exhibit intra-excitonic transitions (such as the energy difference between the 1S and 2P states), leading to a Lorentzian absorption feature in the THz frequency range with a strong negative imaginary contribution that increases with frequency.[49] However, our measurements, shown in Figure S6, reveal a high real component of photoconductivity and only small, positive contributions in the imaginary part, which deviates from the expected response of excitonic behavior. If excitons were present after fast decay, we would expect a long-lived negative imaginary signal, which is absent in our data. Instead, the dynamics are dominated by real conductivity with negligible imaginary contribution. Based on these observations, we exclude exciton formation as the primary relaxation pathway in our sample.

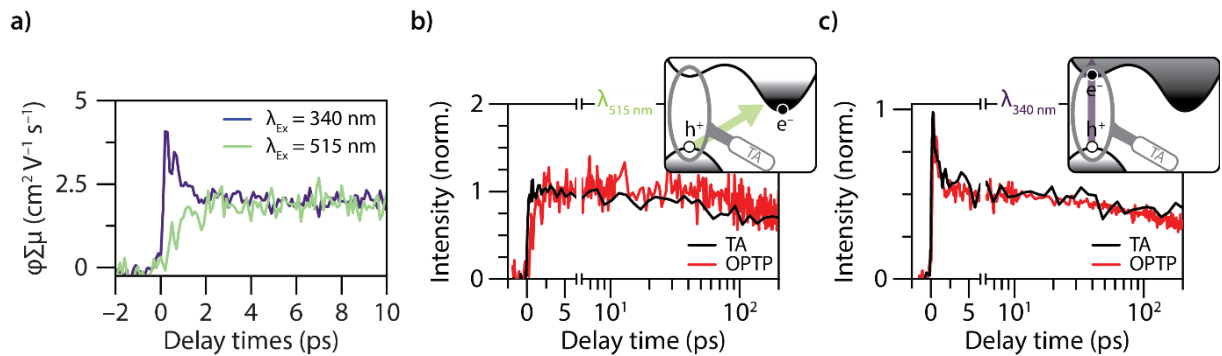

**Figure 4** – Photoconductivity dynamics in $Cs_2AgBiBr_6$ thin films, and its comparison to TA results. **a)** Photoconductivity for bandgap- (515 nm, green) and hot excitation (340 nm, blue). For comparison the TA-decay traces from **Fig. 2c** are reproduced in black and plotted with the mobility traces (red) for **b)** bandgap excitation and **c)** hot excitation. The time traces in **b)** are normalized to unity for the average intensity between 2.5–5 ps. The time traces in **c)** are normalized to unity at the maximum intensities.

With these new insights on the role of charge carrier cooling we establish a full temporal evolution of the charge carrier dynamics in Cs$_2$AgBiBr$_6$ ranging from 0.1 ps to 10 μs (**Fig. 5**). For excitation far beyond the bandgap, highly mobile hot carriers are formed which are rapidly localized upon formation of a polaronic state,[23] at a rate of 0.58 ps$^{-1}$. On basis of bandgap excitation, where both yield and mobility yield product remain constant in the first 10 ps, there is no evidence for cold carrier relaxation into defects on a picosecond timescale. In the subsequent nanoseconds the localized charge carriers can diffuse through the material with an initial mobility of roughly 1 cm$^2$ V$^{-1}$ s$^{-1}$. Previous work showed that on a nanosecond timescale, the diffusion length of the remaining charge carrier population is on the order of the sample's crystalline domains and thickness, ca. 100 nm.[29] As such the significant drop of the photoconductivity compared to the carrier population around 10 ns is attributed to the immobilization of charge carriers at the grain boundaries and/or surface states. The trapped charge carriers remain in the excited-state on a microsecond timescale explaining the long-lived tail in the photoluminescent decay curves (**Fig. S7**),[25] and the increased photoconductance observed for tens of microseconds after the first excitation of double-pulse-excitation TRMC experiments.[50]

The ultrafast hot carrier cooling dynamics in Cs$_2$AgBiBr$_6$ thin films were studied, using multi-resonant pump visible-probe TA experiments. Bandgap vs. hot excitation unambiguously showed that the initial decay kinetics observed in previous pump-probe experiments are the result of hot carrier cooling. Temperature-dependent TA experiments reveal that this cooling rate is independent of temperature and coincides with the temperature-dependent mobility loss observed in previous photoconductivity studies. OPTP experiments on the same sample with the same excitation energies show that the ultrafast apparent photoconductivity loss is the result of cooling of hot carriers to the polaron states. The overlapping photoconductivity decay dynamics 2 ps after photoexcitation, for bandgap and hot excitation, indicate that the same final state with the same yield mobility product is formed, supporting the hypothesis of formation of a polaronic state. In the following nanoseconds the photo-generated charge carriers diffuse through the material towards the crystal's grain boundaries. At these surface states the charge carriers lose mobility, but remain in the excited-state for tens of microseconds. The photoconductance of Cs$_2$AgBiBr$_6$ thin films is considerably reduced upon the formation of the polaronic state. The initial mobility of 2 cm$^2$ V$^{-1}$ s$^{-1}$ over a nanosecond timescale and the energetically separated direct- and indirect transitions make this material an interesting platform to study the formation and dynamics of polaronic states.

**Acknowledgement**

H.J.J and E.M.H. are supported by the Advanced Research Center Chemical Building Blocks (ARC-CBBC). H.J.J., A.-P.M.R, Z.A.V. and S.F. thank the Rowland Institute of Harvard University for generous support. L.G. acknowledges the funding support from the CSC scholarship and Max Planck society.

**Supporting Information**

Experimental methods, fitting procedures and additional figures.

# The Effect of Charge Carrier Cooling on the Ultrafast Carrier Dynamics in $Cs_2AgBiBr_6$ Thin Films

Huygen J. Jöbsis[1,2], Lei Gao[3], Antti-Pekka M. Reponen[2], Zachary A. VanOrman[2], Rick P. P. P. M. Rijpers[1], Hai I. Wang[1,3], Sascha Feldmann[2,4], Eline M. Hutter*[1]

1) Utrecht University, Princetonlaan 8, 3584 CB Utrecht, The Netherlands

2) Rowland Institute, Harvard University, 100 Edwin H Land Blvd, Cambridge, MA 02142, USA

3) Max Planck Institute for Polymer Research, Ackermannweg 10, 55128 Mainz, Germany

4) Institute of Chemical Sciences and Engineering, École Polytechnique Fédérale de Lausanne, Rue de l'industrie 17, 1951 Sion, Switzerland

*Corresponding author: e.m.hutter@uu.nl

# Table of Contents

# Experimental method

**Chemicals.** 99.999% Cesium bromide (CsBr), 99.9% silver bromide (AgBr), 99.998% bismuth bromide (BiBr$_3$), 99.9% anhydrous dimethyl sulfoxide (DMSO), >99.8% iso-propanol (IPA). CsBr, BiBr3, DMSO and IPA were purchased from Merck–Sigma–Aldrich, and AgBr was purchased from Strem. Before use BiBr$_3$ powder was heated at 150 °C for 60 minutes to remove water.

**Synthesis.** In a nitrogen-filled glovebox CsBr, AgBr and BiBr$_3$ were dissolved in DMSO in a 2:1:1 molar ratio resulting in a 0.4 M solution (1.2 mmol CsBr, 0.6 mmol AgBr, and 0.6 mmol BiBr$_3$ in 1.5 mL DMSO). The precursor solution was heated to 100 °C before the spin coating synthesis.

Quartz substrates (15x15x1mm) were cleaned by subsequently sonicating them in demineralized water plus simple washing detergent, acetone and iso-propanol for 20 minutes each. Next, the substrates were put in a Osea UV-ozone cleaner for at least 30 minutes and immediately brought into the glovebox where they were pre-heated at 100 °C for at least 5 minutes before the spin coating method.

Thin layers of Cs2AgBiBr6 were synthesized by depositing 150 µL of the precursor solution on the clean, pre-heated substrate which was subsequently spun at 3000 rpm for 40 s. After 20 s 250 µL anti-solvent (IPA) was deposited on the spinning substrate. For this step the pipet tip was brought as close as possible to the substrate. Finally, the samples were annealed at 280 °C for 5 minutes.

**X-ray diffraction.** The X-ray diffraction pattern was recorded with a Bruker-AXs D8 Phaser in Bragg–Brentano geometry, using a Cu K$\alpha_{1,2}$ = 1.54184 Å, operated at 40 kV and 40 mA. For the experiment a step size of 0.02° and a scan speed of 1 s, with a 2 nm slit of the source was used.

**Steady state optical properties.** The transmittance, *T*, absorptance, *A*, and reflectance, *R*, were determined using a Perkin–Elmer lambda UV–vis–NIR lambda950S equipped with an integrating sphere. To record *R* the sample was placed at the back of the integrating sphere under an angle of 8° with respect to the incident light. Next, the sample was placed the center of the integrating sphere under an angle of 10° to record *R+T*. Assuming that that the sum of the *A*, *R* and *T* equals to the total incident light (so ignoring scattering) the *A* and *T* can be determined following:

**Equation S1 a, b:**

$$A = 100\% - (R + T)$$

and

$$T = 100\% - A - R$$

Temperature-dependent UV–vis experiments were performed in a nitrogen-cooled Oxford Instrument Optistat DN cryostat in a EOS multichannel pump probe transient absorption spectrometer (Ultrafast Systems LLC.). The probe pulse (200–2400 nm) is split with a 50/50 beam splitter where half of the light is used as a reference and the other half is directed towards the sample to record $T$. The sample was cooled from 300–80 K in steps of 5 K. The room temperature $R$ was used to determine $A$.

**Fs-transient absorption spectroscopy.** Femtosecond transient absorption measurements were performed using a setup based on modules supplied by Light Conversion, with a 1030 nm seed laser (PHAROS, Light Conversion, Yb:KGW lasing medium, 400 µJ pulse energy, 200 fs duration, operated at 5 kHz repetition rate). The 515 nm or 343 nm pump beam (ca. 900 µm in diameter) was generated from the seed using a harmonic generation unit (HIRO, Light Conversion) via non-linear crystals (BBO, lithium triborate). Two probe beam (ca. 200 µm in diameter) were generated from the seed laser or second harmonic of the fundamental using supercontinuum generation in a sapphire crystal. The probe generated from the 1030 nm fundamental spanned from ca. 540–960 nm. The probe generated from the 515 nm second harmonic spanned from ca. 375–500 nm. The pump-probe delay was controlled over a range of 7 ns by changing probe path length via a multipass delay stage, and the pump was passed through an optical chopper (100 Hz), where a beamsplitter/photodiode combination was used to divide and sort measurements into pumped and unpumped. The probe beam was passed into a grating spectrograph (Andor Kymera 193i, grating blaze wavelength 800 nm) and recorded using a Si NMOS photodiode array detector (256 pixels).

**Chirp correction.** The probe pulse arrival time has a wavelength dependence (especially when measuring in the cryostat), which is corrected for by applying a wavelength-dependent time offset determined by a third-order polynomial fit to selected points in the coherent artifact at time zero.

**Optical-pump-terahertz probe spectroscopy.** We operate the OPTP spectrometer using a commercial mode-locked Ti:sapphire fs-laser with 1 kHz repetition rate. The pulse duration is approximately 100 fs, and the central wavelength is around 800 nm. A single-cycle terahertz pulse is generated through optical rectification on a 1 mm thick ZnTe crystal. The time-domain terahertz electric field is measured via the electro-optic sampling method by a second ZnTe crystal. For selective optical excitations, the wavelengths are produced by a commercial optical parametric amplifier from Light Conversion.

## Additional Figures

**Figures S1 – Characterization of Cs2AgBiBr6 thin film. a)** X-ray diffraction pattern of Cs2AgBiBr6 thin film plotted as a function of scattering vector *q*. The Rietveld refinement fit (black) confirms the elpasolite crystal structure and reveals a small presence of unreacted AgBr. The green ticks indicate the Bragg-reflection positions of $Cs_2AgBiBr_6$ (top) and AgBr (bottom). The quartz substrate causes the broad background around 15 nm⁻¹. The flat residual (grey) indicates that no other crystalline phases are present. **b)** The layer thickness, *L*, is approximated by considering that $T = e^{-\alpha L}$ with *T* the transmittance and $\alpha$ the absorption coefficient which was previously reported by us.[1]

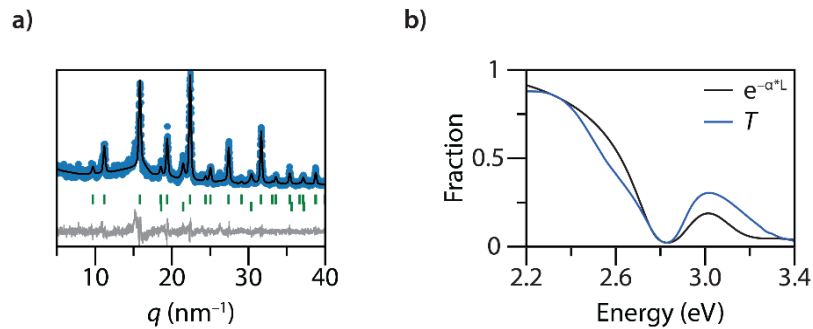

**Figure S2 – Fluence-dependent bleach area and position.** The excitation density was determined as described in Supplementary Note **S1**. **a–d)** The dashed line is a power law, $y = a * x^p$, fitted to the experimental data with $p \approx 1$ for $t = 1$ ps and $t = 5$ ns. For $t = 100$ fs the exponent $p < 1$ indicates higher order decay dynamics influence the TA spectra. **e, f)** The time dependent bleach maximum for different excitation densities. The bleach centered around 2.8 eV is fitted with a gaussian distribution to track the bleach position over time. **g,h)** Spectral $\Delta T/T$ maps for beyond-bandgap and sub-bandgap excitation energies, respectively, at similar excitation densities. **i,j)** Spectral slices at various delay times for beyond-bandgap and sub-bandgap excitation, respectively.

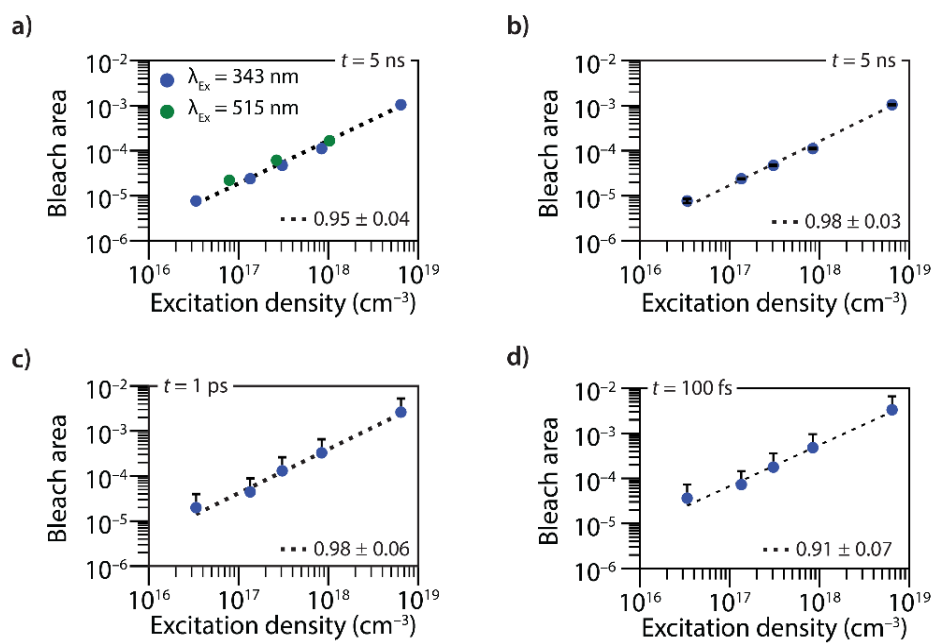

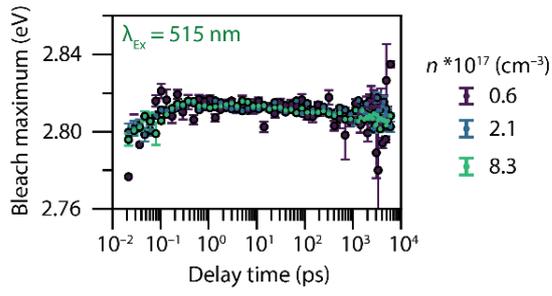
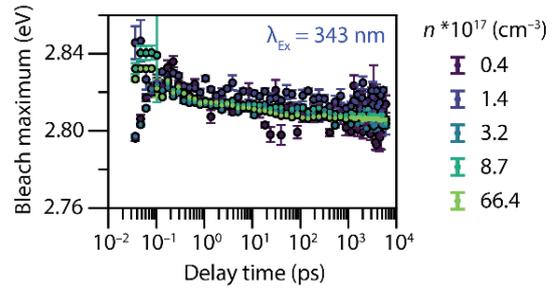
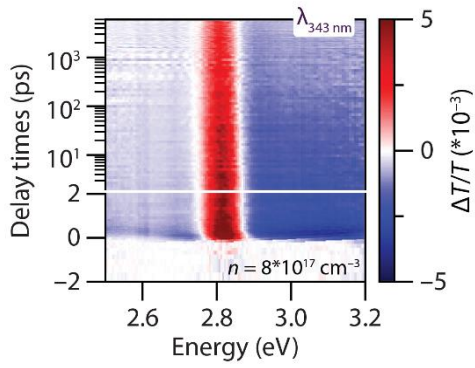
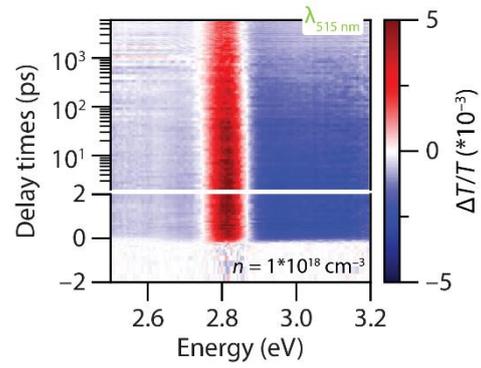
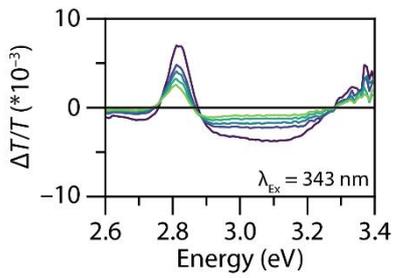
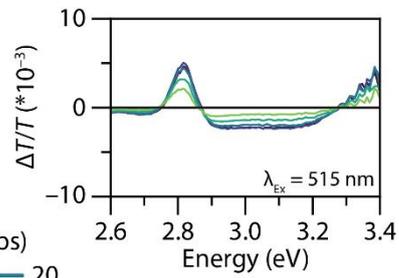

**Figure S3 — Effect of fitting region on extracted TC. a–c)** Normalized ΔT/T spectra for 0.1–10 ps delay times (colored) and the Maxwell–Boltzmann distribution function fitted to different lengths of the high-energy tail of the bleach feature (red). **d)** The extracted values for TC for the different fitting regions in **a–c)**. **e–g)** Normalized ΔT/T spectra for 0.1–10 ps delay times (colored) and the Maxwell–Boltzmann distribution function fitted to different heights of the normalized bleach feature (red). **h)** The extracted values for TC for the different fitting regions in **e–g)**.

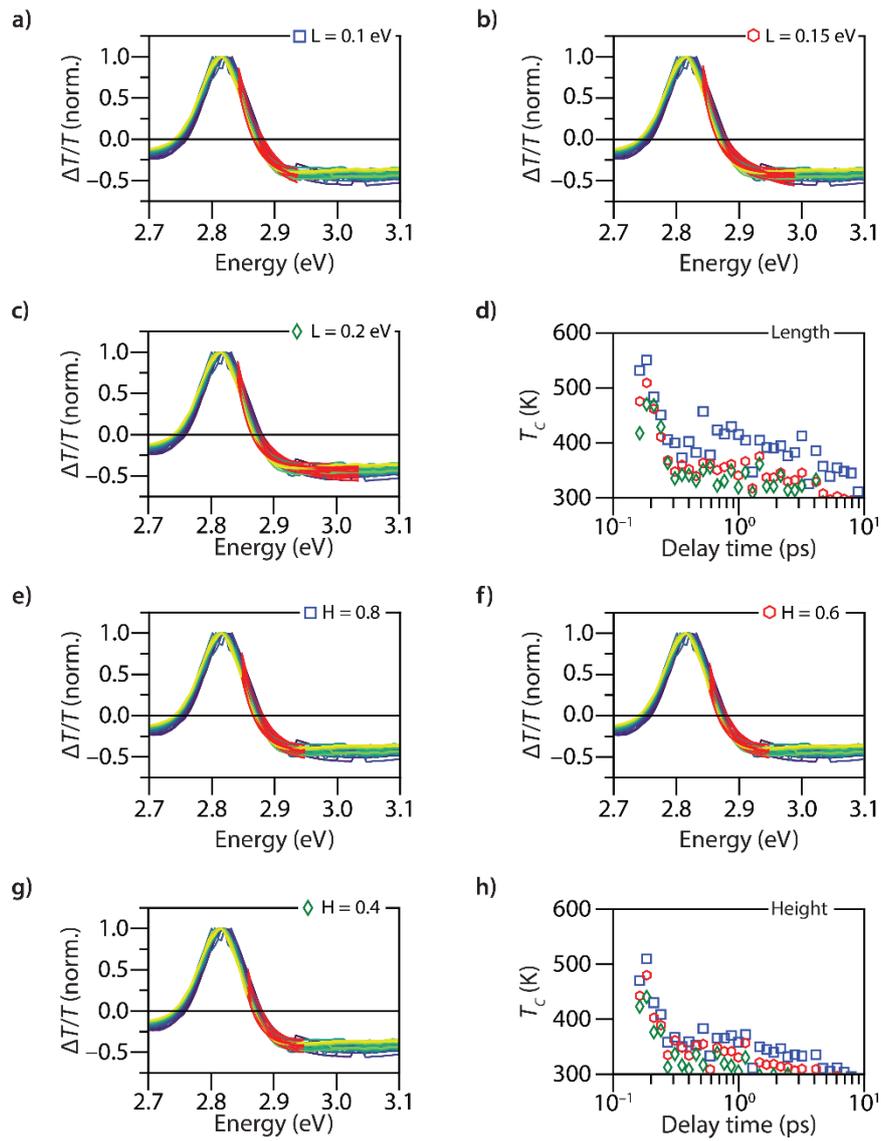

**Figure S4 – Temperature dependent UV–vis of Cs$_2$AgBiBr$_6$. a)** The absorption onset was defined as the point where 10% of the incident probe light is absorbed by the sample. The exciton peak position was determined by fitting a Gaussian distribution to the peak between 2.78 eV and 2.91 eV. **b)** Zoom in of the absorption onset when cooling down (blue) and heating up (red) showing a slight hysteresis around 100 K which is in line with a previously reported phase transition from cubic to tetragonal (when cooling down) around this temperature.[2]

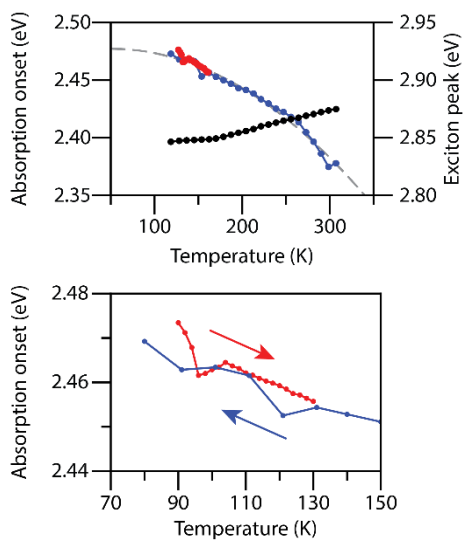

**Figure S5 – Temperature-dependent TA spectroscopy.** Ultrafast recovery traces at 2.82 eV when exciting at 343 nm and 515 nm in- and outside the cryostat at room temperature (ca. 300 K). The additional chirp due to the windows of the cryostat lowers the time resolution of the experiments performed in the cryostat as can be seen by the truncated peak of the decay trace.

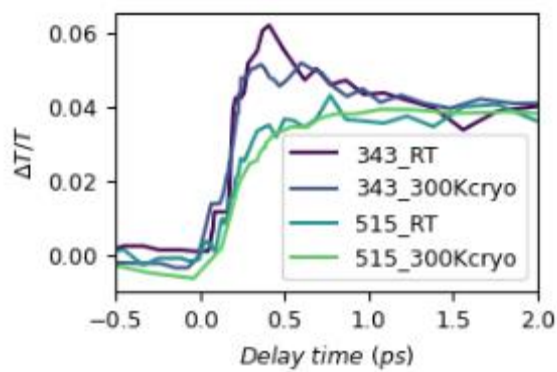

**Figure S6 — THz spectroscopy experiments.** In the left panel the photoconductivity, σ, divided by the photon fluence, is plotted as a function of delay times for bandgap (green) and hot (blue) excitation. On the right the complex photoconductivity is plotted as a function of delay time.

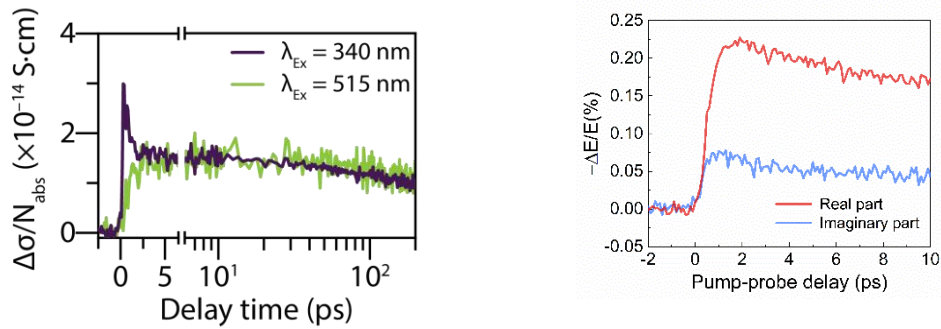

**Figure S7 — Time-resolved photoluminescence spectroscopy. a)** Steady-state photoluminescence (PL), absorptance, transmittance and reflectance as a function of photon energy. **b)** The time-resolved PL was recorded using an inhouse build photoluminescence setup equipped with an APD detector and a 650 nm (1.9 eV) bandpass filter.

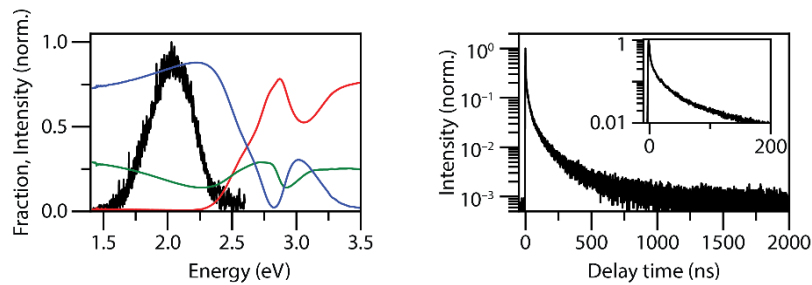

For OPTP spectroscopy, we fixed the sampling pulse at a specific point in the THz waveform and then record its change as a function of pump-probe delay time t. This way, we obtain the time-resolved THz transmission change which is related to the carrier dynamics. When the sampling pulse is fixed at the peak of the main THz pulse, we measure the pump-induced intensity change, which is related to the real part of photoconductivity. When the sampling pulse is fixed at the zero-crossing point, we get the imaginary part of photoconductivity. In the close vicinity of this crossing point, the phase modulation (in the $t_S$ axis) is linearly proportional to the amplitude modulate in the Y (electrical field) axis.

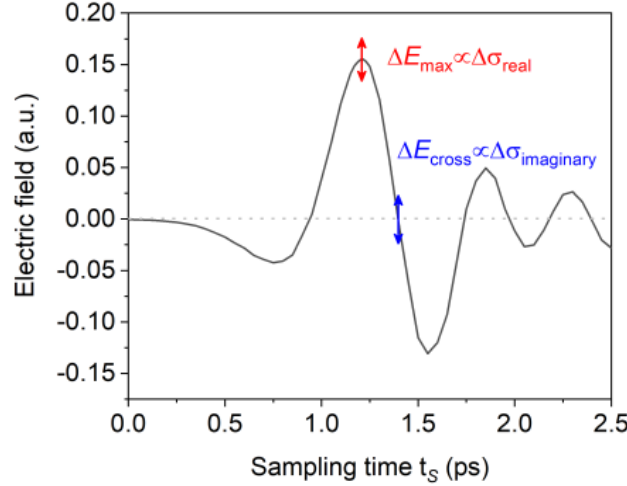

**Figure S8 — Real and imaginary photoconductivity measurement in 1D OPTP configuration.** When the sampling pulse is fixed at the THz peak (red arrow), the photoinduced modulation of the THz transmission reports the real part of the (frequency-average) photoconductivity, while the imaginary part of the photoconductivity is measured the pump-induced THz field changes at the zero-crossing (blue arrow).

**Supplementary Note 1 — Estimating the carrier density**

The initial photo-excited carrier density is calculated following:

**Equation S3**

$$n = \frac{J * A}{d}$$

Where *J* is the pump fluence (photons per $cm^{-2}$ per pulse), *A* the absorptance and *d* the sample thickness. We note, however, that the penetration depth of the 343 nm pump is limited to the first 10 nm of the sample, whereas the penetration depth for the 515 nm pump is larger than the sample thickness (Fig. S1). The pump fluence is determined by recording the pump power, *P*, just before the sample and the area of the pump at the sample position, using:

**Equation S4**

$$J = \frac{P}{h\nu * f * area}$$

With *hv* the photon energy and *f* the pump frequency. As a rule of thumb the pump spot size should be larger than the probe spot to ensure a homogeneous excitation in the probed region. Here, the pump spot diameter was ca. 1000 μm and the probe spot ca. 400 μm.

**Supplementary Note 2 — Extracting the carrier temperature**

With the use of transient absorption (TA) spectroscopy the photo-induced carrier temperature, $T_C$, can be approximated.[3,4] For this a quasi- thermal equilibrium is assumed, meaning that there is no net exchange of charge carriers between different points in the semiconductor and/or between the semiconductor and its surrounding. Under such equilibrium conditions, Fermi–Dirac statistics describe the energy distribution in the material following:

**Equation S5**

$$f_0(E, E_F, T) = \frac{1}{1 + e^{E-E_F/k_B T}}$$

with $E_F$ the average electrochemical potential of electrons in the semiconductor, also known as the Fermi energy or Fermi level. Upon hot excitation, i.e. $E \gg E_F$, Eq. S5 can be approximated by a Maxwell–Boltzmann distribution function that represents the probability of an electron occupying an energy state well above $E_F$ with an average carrier temperature $T_C$:

**Equation S6**

$$f(E \gg E_F, T) \approx e^{-E/k_B T}$$

In TA spectroscopy the intensity of the difference spectrum scales with the excited state population. At the resonance frequency of an optical transition (e.g. bandgap excitation) reduced absorption (increased transmission) due to the occupation of excited states results in a negative $\Delta A$ (positive $\Delta T/T$) gaussian shaped signal. For pump energies well above the bandgap, hot carriers populate energy states well above $E_F$ with a temperature $T_C$ that is higher than the lattice/ambient temperature. As a consequence the bleach signal asymmetrically broadens on the high-energy side (here 2.85–2.92 eV) of the feature. As such by fitting the high energy side of the TA spectrum $T_C$ can be approximated using:

**Equation S7**

$$\frac{\Delta T}{T} = A * e^{-E/k_B T_C} + background$$

with the fitting parameters A, an amplitude that depends on experimental parameters like sample thickness and excitation density, $T_C$ the carrier temperature, and a background to fit the negative $\Delta T/T$ signal. In the literature it is pointed out that this technique depends on a arbitrarily chosen fit region to determine $T_C$. In figure S3 we, however, show that for different fitting regions we obtain similar values for $T_C$ and a similar cooling rate. Justifying the use of equation 5 to determine $T_C$ for this material.